\documentclass[prl,twocolumn,showpacs]{revtex4}  
\usepackage{graphicx}  
\usepackage{amssymb}   

\begin{document}


\title{Stimulated Emission of High-Frequency Phonons
\footnote{Sov. Phys. Solid State, 1978, Vol. 20, No.10, pp. 1807-1808.} }

\author{G.\,V.\, Kovalev \/\thanks}



\affiliation{Engineering-Physics Institute, Moscow }


\date{Apr. 24, 1978}

\begin{abstract}
The motion of channeled particles is accompanied by the phonon emission. This feature can be used for the stimulated  generation of phonons with large wave vectors and  frequencies. 
\end{abstract}

\maketitle


Difficulties are encountered [1] when the usual methods for the excitation of phonons are extended to  large wave vectors ${\bf q} \approx 0.5 \cdot 10^5 \, cm^{-1}$ . For example, when such phonons are generated by the Raman scattering of laser 
light, it is necessary to use lasers emitting in the ultraviolet range. We shall  consider the generation of phonons with ${\bf q \approx  K}$ ($\bf{K}$ is the reciprocal lattice vector) by  a fast charged particle moving in a crystal under channeling conditions. The motion of channeled particles is known to be characterized by discrete transverse oscillations and free 
motion in the longitudinal direction. The interaction with  thermal vibrations of the lattice facilitates transitions of a particle between discrete transverse levels accompanied  
by the emission of several phonons. If the gap between the levels $\Delta \, \epsilon_{n n'}$ is less than the Debye frequency $w_{D}$, the transition of a particle from a level $n$ to a level $n'$ characterized by conservation of the longitudinal momentum is accompanied mainly by one-phonon emission when the energy of new phonons $\omega_k$ is $\Delta \, \epsilon_{n n'}$ in the transverse 
direction. The value of $\Delta \, \epsilon_{n n'}$ may vary within wide limits 
with the energy of the particle. In the case of planar 
channeling of positrons in a parabolic potential [2] we have 
$\Delta \, \epsilon_{n n'} \approx 10 \gamma^{-1/2}$ eV   [$\gamma = (1-v^2)^{-1/2}$ is the relativistic factor and $\hbar=c=1$].

The matrix element of a transition of a particle from 
an initial state $ \Psi_{i} = \exp[i {\bf p_{\parallel
} r_{\parallel } }] u_n (x)$ to a final state  $ \Psi_{f} = \exp[i {\bf p'_{\parallel } r_{\parallel } }] u_{n'} (x) $ accompanied by a transition of a crystal 
from a state $ \Phi_{\beta} $ to a  state $ \Phi_{\beta'} $ is 
\begin{eqnarray}
M^{f i} _{\beta' \beta}=\left\langle \Phi_{\beta'} \Psi_{f} | W  | \Psi_{i}  \Phi_{\beta} \right\rangle, 
\label{r1}
\end{eqnarray}
where $W({\bf  r, \{ R_j \} }) $ is the difference between the true potential  $V({\bf  r, \{ R_j \} }) =\sum_{j, {\bf q}} V_{q} \exp i {\bf q (r-R^{0}_j- u_j})$ and the potential, 
averaged along the $y O z$ plane, of atoms at rest at the 
equilibrium positions $\bar{U} ({\bf  r, \{ R^{0}_j \} })$ [3].   
For a purely  inelastic process we have $\left\langle \Phi_{\beta'}| \bar{U} |  \Phi_{\beta} \right\rangle=0$, and, consequently, the nondiagonal terms $M^{f i} _{\beta' \beta}$ are governed only by the potential $V({\bf  r, \{ R_j \} }) $.  An explicit calculation of the matrix element (1) for the process involving $s$ phonons 
gives 
\begin{eqnarray}
M^{f i} _{\beta' \beta}=\sum_{j, {\bf q}} e^{-Z_j /2}  \delta_{ {\bf p_{\parallel}-p'_{\parallel}, q_{\parallel} }} V_{q} \, e^{- i {\bf R^0_j (q- \sum_{\alpha = 1}^{s} \pm f_{\alpha})}} \times \nonumber \\
\times \int{u^{*}_{n'} u_{n} e^{i q_x x}   d x} \prod^{s}_{\alpha =1 }{({\bf q, e_j(\beta_{\alpha})}) \sqrt{ \frac{n_{\beta \alpha} + \frac{1}{2} \pm \frac{1}{2} }{2 M_j n \omega (\beta_{\alpha})}} },
\label{r2}
\end{eqnarray}
where 
\begin{eqnarray}
Z_{j} =\frac{1}{2 M_j  N } \sum_{\beta} \frac{( {\bf q, e_j(\beta)} )^{2} }{\omega(\beta) } (2 n_{\beta}+1), \nonumber
\label{ra}
\end{eqnarray}
and  $\beta_{\alpha} = ({\bf  f_{\alpha}, \xi_{\alpha}} )$  is the combination of the wave vector with the branch number of a phonon $\alpha$. The following laws of 
conservation of momentum and energy are satisfied: 
\begin{eqnarray}
 {\bf p_{\parallel}-p'_{\parallel}} = \sum^{s}_{\alpha =1 } { \pm {\bf f_{\parallel  \alpha}+ K_{\parallel}}}, \nonumber \\
E - E'  =   \sum^{s}_{\alpha =1 } { \pm  \omega_{\alpha} }. 
\label{r3}
\end{eqnarray}

We shall consider phonon generation in the case when a particle undergoes a transition from one level to another 
($\Delta \, \epsilon_{n n'} \neq 0$ , where $\Delta \, \epsilon_{n n}=0$  corresponds to the usual 
Cherenkov phonon emission). In the relativistic case it is found from Eq. (3) that 
\begin{eqnarray}
 \sum^{s}_{\alpha =1 } { \pm  \omega_{\alpha} } \cong \Delta \, \epsilon_{n n'} +  \sum^{s}_{\alpha =1 }{ \pm { f_{\parallel  \alpha}+ K_{\parallel}}}.
\label{r4}
\end{eqnarray}
In the scattering process accompanied by the emission of 
one phonon we can confine our attention to the $K_{\parallel}$ case [if 
 $K_{\parallel} \neq 0$ , the right-hand side of Eq. (4) is $\approx 10^3 eV  >> \omega_D$]. 
Using Eqs. (2) and (4), we obtain easily the probability of 
phonon emission in an angle $d \Omega_{k}$: 
\begin{eqnarray}
\frac{d w}{d \Omega_{k}}=N  \sum_{{\bf  K_{\perp}} }  \exp(- \bar{Z})  V^2_{{\bf  f_k + K_{\perp} } } \frac{({\bf  f_k + K_{\perp} , e})^2}{2 M} \nonumber \\
\times 
 \frac{|\left\langle n' | e^{i(f_{k \perp} + K_{\perp})x}  | n \right\rangle|^2}{\epsilon_{n n'} + f_{k \parallel}} (\bar{n}_{k}+1).
\label{r5}
\end{eqnarray}

It is clear from this expression that for $\Delta \, \epsilon_{n n'} > 0$ the phonon  emission is mainly along the direction $\theta \sim \pi$ and the 
phonon frequency is  $\omega_k \sim \Delta \, \epsilon_{n n'} v_s$, where $v_s$ is the velocity 
of sound; these phonons have the transverse polarization. 

If  $\Delta \, \epsilon_{n n'} < 0$ (transition to an upper level), the intensity 
maximum corresponds to $\theta \sim 0$, and the phonon frequency 
is then $\omega_k \sim - \Delta \, \epsilon_{n n'} v_s$.  In the Cherenkov emission case 
($\Delta \, \epsilon_{n n'}= 0$) the direction is given by $\theta \sim \pi/2-v_s$. The 
first two cases can be used to determine the energy of a 
particle, because $\Delta \, \epsilon_{n n'}= f(E)$.

Moreover, this feature can be used for the stimulated 
generation of phonons with large wave vectors and of frequency  $w_D$. This can be done by establishing an inverted population of the particles in the channel 
\begin{eqnarray}
\Delta N=  j \, S \frac{l}{v}(a_n^2- a^2_{n'} ), 
\label{r6}
\end{eqnarray}
where $j$ is the density of the current, $S$ is the beam cross 
section, $l$ is the thickness of the crystal, $v$ is the particle 
velocity, and $a$ is the coefficient of matching of a plane 
wave outside a crystal with the wave functions of the 
particle inside the crystal. The threshold population inversion should satisfy the following simple condition [4] 
\begin{eqnarray}
\Delta N_{th} =   \frac{\rho}{\tau_c w_{n n'}}, 
\label{r7}
\end{eqnarray}

where $w_{ n'n}$  is the probability of a transition of a particle 
from a state $n$ to a state $n'$, $\tau_c$ is the phonon lifetime in the 
crystal, and $\rho$ is the number of different of vibration modes 
in a phonon spectral line. The threshold current density 
is obtained from Eqs. (6) and (7): 
\begin{eqnarray}
j_{th} =\frac{\rho}{\tau_c w_{n n'}}  \frac{v}{l S (a_n^2- a^2_{n'} )}. 
\label{r8}
\end{eqnarray}
For the parameters $E = 100 $MeV, $\tau_c = 10^{-9}$ sec, $S = 1$ $mm^2$ , $l = 1 \mu$,  $a_n^2- a^2_{n'} = 10^{-3}$,  and $\rho = 10^6$, we find that $j _{th} =10^{18}$ particles $\cdot  cm^{- 2} \cdot sec^{-1}$.

The author is grateful to N. P. Kalashnikov and M. N.  Strikhanov for a discussion of the results obtained.


\end{document}